\documentclass{PoS}
\pdfoutput=1
\usepackage{amsmath}
\usepackage{mathbbol}
\usepackage{placeins}
\usepackage{braket}

\usepackage{feynmp}
\ifpdf
\fi

\title{Potentials between pairs of static-light mesons}
\ShortTitle{Potentials between pairs of static-light mesons}

\author{Gunnar Bali, \speaker{Martin Hetzenegger}\\
		Institut f\"ur Theoretische Physik, Universit\"at Regensburg,\\ 
		93040 Regensburg, Germany\\
		E-mail:
		\email{gunnar.bali@physik.uni-regensburg.de},
		\email{martin.hetzenegger@physik.uni-regensburg.de}\\\vspace*{.25cm}}
\author{
\rm\centering (QCDSF Collaboration)}

\abstract{We give an update on our ongoing investigations of potentials between pairs of static-light mesons, ${\mathcal B}({\mathbf r}){\mathcal B}({\mathbf 0})$ and ${\mathcal B}({\mathbf r})\overline{\mathcal B}({\mathbf 0})$, in $N_{\mathrm f}=2$ Lattice QCD, in different spin and isospin channels. The question of attraction and repulsion is particularly interesting with respect to the $X(3872)$ charmonium state and charged candidates such as the $Z^{+}(4430)$. We employ the nonperturbatively improved Sheikholeslami-Wohlert fermion and the Wilson gauge actions at two lattice spacings $a\approx 0.084$ fm and $a\approx 0.077$ fm with a pseudoscalar mass of $m_{\mathrm PS}\approx 770$ MeV and $m_{\mathrm PS}\approx 400$ MeV respectively. We use stochastic all-to-all propagator techniques, improved by a hopping parameter expansion. The analysis is based on the variational method, utilizing various source and sink interpolators.}

\FullConference{ The XXIX International Symposium on Lattice Field Theory - Lattice 2011\\
July 10-16, 2011\\
Squaw Valley, Lake Tahoe, California}

\begin{document}
\section{Introduction}
Potentials between static-light mesons ($\mathcal{B}=Q\bar{q}$) are of interest since they give insights in the nature of strong interactions from first principles for multiquark systems. For large heavy quark masses, e.g., the spectra of heavy-light mesons are determined by excitations of the light quark and gluonic degrees of freedom. In particular, the vector-pseudoscalar splitting vanishes and the static-light meson $\mathcal{B}$ can be interpreted as either a $\overline{B}$, a $\overline{B}^*$, a $D$ or a $D^*$ heavy-light meson. Calculating potentials between two $\mathcal{B}$ mesons then will also enable investigations of possible bound tetraquark states or for particles that are close to the meson-antimeson thershold, such as the $X(3872)$ or the $Z^{+}(4430)$. 
Therefore, $\mathcal{BB}$ and $\mathcal{B\overline{B}}$ potentials have been studied by many groups.
First calculations were performed by Michael and Pennanen~\cite{Michael:1999nq,Pennanen:1999xi}. A more detailed quenched study can be found in ref.~\cite{Detmold:2007wk}. For the computation of static-light meson-antimeson systems we refer to~\cite{Sesam}. Recent dynamical simulations with twisted mass fermions were carried out by Wagner in refs.~\cite{Wagner:2010ad,Wagner:2011ev} and with Sheikholeslami-Wohlert fermions in our Lattice 2010 proceedings~\cite{Bali:2010xa}.

\section{Computation}
In these proceedings we present the latest results from our investigations of potentials between two static-light mesons. Of special interest is the question of attraction and repulsion and their dependence on the separation between the static quarks. Therefore, we numerically determine ground and excited states of ${\mathcal B}$ mesons as well as intermeson potentials between pairs of static-light mesons, ${\mathcal B}({\mathbf r}){\mathcal B}({\mathbf 0})$ and ${\mathcal B}({\mathbf r})\overline{\mathcal B}({\mathbf 0})$. The static quark-quark (or quark-antiquark) separation is given by $r=|{\mathbf r}|=Ra, R\in \mathbb{N}_0$. $a$ denotes the lattice spacing, $Q$ a static colour source and the positions of the mass-degenerate light quarks $q \in\{u,d\}$ are not fixed. Thus, ${\mathcal B}$ mesons carry isospin $I=1/2$ and for $\mathcal{BB}$ and $\mathcal{B\overline{B}}$ states $I \in \{0,1\}$ with $I_z\in \{-1,0,1\}$. For ${\mathcal B}$ mesons the isosinglet corresponds to the representation $q_1 q_2 = ud + du$ and the isotriplet to $q_1 q_2 \in \{uu, dd, ud-du\}$ respectively. The corresponding representations for $\mathcal{B\overline{B}}$ states are given by $\bar{q}_1 q_2 = \bar{u}u + \bar{d}d$ for $I=0$ and $\bar{q}_1 q_2 \in \{\bar{d}u, \bar{u}d, \bar{u}u-\bar{d}d\}$ for $I=1$. Graphically, the quark line diagrams with respect to isospin that we evaluate can be depicted as,
\begin{align}
\mbox{Isospin $I=0$:} & \quad C_{\mathcal{BB}}(t) = \;\;
\parbox{.06\textwidth}{
\begin{fmffile}{diag_para}
\begin{fmfgraph*}(20,27)
\fmfstraight
\fmfkeep{static-light}
\fmfleft{lu,lo}
\fmfright{ru,ro}
\fmfbottom{lu,um,ru}
\fmf{fermion}{lu,lo}
\fmf{fermion}{ru,ro}
\fmf{wiggly,tension=0,left=0.5}{lo,lu}
\fmf{wiggly,tension=0,right=0.5}{ro,ru}
\end{fmfgraph*}
\end{fmffile}},
 &&
C_{\mathcal{B\overline{B}}}(t)  = \; \;
\parbox{.06\textwidth}{
\begin{fmffile}{diag_antipara}
\begin{fmfgraph*}(20,27)
\fmfstraight
\fmfkeep{static-light}
\fmfleft{lu,lo}
\fmfright{ru,ro}
\fmfbottom{lu,um,ru}
\fmf{fermion}{lu,lo}
\fmf{fermion}{ro,ru}
\fmf{wiggly,tension=0,left=0.5}{lo,lu}
\fmf{wiggly,tension=0,left=0.5}{ru,ro}
\end{fmfgraph*}
\end{fmffile}}
-\;2\;\;
\parbox{.06\textwidth}{
\begin{fmffile}{diag_box}
\begin{fmfgraph*}(20,27)
\fmfstraight
\fmfkeep{static-light}
\fmfleft{lu,lo}
\fmfright{ru,ro}
\fmf{fermion}{lu,lo}
\fmf{fermion}{ro,ru}
\fmf{wiggly}{lu,ru}
\fmf{wiggly}{lo,ro}
\end{fmfgraph*}
\end{fmffile}}, \\ 
\mbox{Isospin $I=1$:} & \quad C_{\mathcal{BB}}(t) = \;\;
\parbox{.06\textwidth}{
\begin{fmffile}{diag_para}
\begin{fmfgraph*}(20,27)
\fmfstraight
\fmfkeep{static-light}
\fmfleft{lu,lo}
\fmfright{ru,ro}
\fmfbottom{lu,um,ru}
\fmf{fermion}{lu,lo}
\fmf{fermion}{ru,ro}
\fmf{wiggly,tension=0,left=0.5}{lo,lu}
\fmf{wiggly,tension=0,right=0.5}{ro,ru}
\end{fmfgraph*}
\end{fmffile}}
-\;\;
\parbox{.06\textwidth}{
\begin{fmffile}{diag_cross}
\begin{fmfgraph*}(20,27)
\fmfstraight
\fmfkeep{static-light}
\fmfleft{lu,lo}
\fmfright{ru,ro}
\fmf{fermion}{lu,lo}
\fmf{fermion}{ru,ro}
\fmf{wiggly}{lo,ru}
\fmf{wiggly}{lu,ro}
\end{fmfgraph*}
\end{fmffile}},
 &&
C_{\mathcal{B\overline{B}}}(t)  = \; \;
\parbox{.06\textwidth}{
\begin{fmffile}{diag_antipara}
\begin{fmfgraph*}(20,27)
\fmfstraight
\fmfkeep{static-light}
\fmfleft{lu,lo}
\fmfright{ru,ro}
\fmfbottom{lu,um,ru}
\fmf{fermion}{lu,lo}
\fmf{fermion}{ro,ru}
\fmf{wiggly,tension=0,left=0.5}{lo,lu}
\fmf{wiggly,tension=0,left=0.5}{ru,ro}
\end{fmfgraph*}
\end{fmffile}} .
\end{align}
Straight lines represent static quark propagators and wiggly lines light quark propagators.
\begin{table}[b]
\centering
\begin{tabular}{|c|l|l|l|l|l|l|l|}
	\hline
	volume $L_{\sigma}^3\times L_{\tau}$   & $\beta$ & $\kappa_{\mathrm{val}} = \kappa_{\mathrm{sea}}$  & $c_{\mathrm{SW}}$ & $a/\mathrm{fm}$ & $La/\mathrm{fm}$ & $m_{\mathrm{PS}}/\mathrm{MeV}]$ & $N_{\mathrm{conf}}$ \\
	\hline
	$16^3\times32$ & $5.29$  & $0.13550$ & $1.9192$ & $0.084$  & $1.34$  & $770(9)$ & $180$ \\
	$24^3\times48$ & $5.29$  & $0.13620$ & $1.9192$ & $0.077$  & $1.85$  & $400(4)$ & $200$ \\
	\hline  
\end{tabular}
\caption{Lattice parameters.}
\label{tab:lattice}
\end{table}

Masses are calculated from the asymptotic behavior of Euclidean-time correlation functions. We employ $N_{\mathrm f}=2$ Sheikholeslami-Wohlert configurations generated by the QCDSF Collaboration~\cite{AliKhan:2003br}. The parameter values are listed in table~\ref{tab:lattice}, where the scale is set using $r_0(\beta,\kappa)=0.5$~fm. The pseudoscalar mass corresponds to its infinite volume value. We use the Chroma software system~\cite{Edwards:2004sx}. For the techniques and improvement methods we use, we refer to~\cite{Bali:2010xa} and the references therein. To analyze our data and to extract also excited states we apply the variational method~\cite{Vari}, solving a generalized eigenvalue problem for a $3\times 3$ cross correlation matrix generated by different amounts of Wuppertal smearing~\cite{Wuppertalsmear} applied to the source and sink operators. Errors are calculated using the jackknife method.

\section{Representations and classification of states}
To create $\mathcal{B}$ meson states as well as $\mathcal{BB}$ and $\mathcal{B\overline{B}}$ systems of different $J^{P(C)}$ we use interpolators $\mathcal{B}=Q\mathcal{O}\bar{q}$, where the operators $\mathcal{O}$ contain combinations of Dirac $\gamma$-matrices and covariant lattice derivatives. This has been discussed in~\cite{Bali:2010xa} and we give a summary for the representation and classification of our states.\\
In the continuum limit, the static-light states can be classified according to fermionic representations $J^P$ of the rotation group ${\mathrm O(3)}$. At vanishing distance ${\mathbf r}={\mathbf 0}$ the ${\mathcal B}{\mathcal B}$ and ${\mathcal B}\overline{\mathcal B}$ states can be characterized by integer $J^P$ and $J^{PC}$ quantum numbers, respectively. However at $r=|{\mathbf r}|>0$ the $\mathrm{O(3)}$ (or $\mathrm{O(3)}\otimes{\mathcal C}$) symmetry is broken down to its cylindrical $\mathrm{D_{\infty h}}$ subgroup. The irreducible representations of this are conventionally labeled by the spin along the axis $\Lambda$, where $\Sigma,\Pi,\Delta$ refer to $\Lambda=0,1,2$, respectively, with a subscript $\eta=g$ for gerade (even) $PC=+$ or $\eta=u$ for ungerade (odd) $PC=-$ transformation properties with respect to the midpoint. All $\Lambda\geq 1$ representations are two-dimensional. The one-dimensional $\Sigma$ representations carry an additional $\sigma_v=\pm$ superscript for their reflection symmetry with respect to a plane that includes the two endpoints.
\begin{table}[b]
\centering
\begin{tabular}{|c|c|c|c|c|}
	\hline
	\textbf{$\mathcal{O}$} & wave~\protect\cite{Michael:1998sg} & $\mathrm{O_h}'$ rep. &continuum $J^{P}$ & $J^{P}$ (heavy-light)\\
\hline
$\gamma_5$&$S$&$G_1^{+}$&$\frac{1}{2}^{+}$&$ 0^{-}, 1^{-}$\\
$\Eins$&$P_-$&$G_1^{-}$&$\frac{1}{2}^{-}$&$ 0^{+}, 1^{+}$\\
$\gamma_i\nabla_i$&$P_-$&$G_1^{-}$&$\frac{1}{2}^{-}$&$0^{+}, 1^{+}$\\
$\left( \gamma_1\nabla_1 - \gamma_2\nabla_2 \right) + \mathrm{cycl.}$&$P_+$& $H^{-}$   & $\frac{3}{2}^{-}$& $ 1^{+}, 2^{+}$  \\
	\hline  
\end{tabular}
\caption{Operators and representations for static-light mesons. In the last column we display the $J^P$ for a heavy-light meson, obtained by substituting the (spinless) static source by a heavy fermion.}
\label{tab:operators_sl}
\end{table}

The operators that we used to create the static-light mesons are displayed in table \ref{tab:operators_sl}. The intermeson potentials were obtained by combining two static-light mesons of different (or the same) quantum numbers. This can be projected into an irreducible $\mathrm{D_{\infty h}}$ representation, either by coupling the light quarks together in spinor space~\cite{Wagner:2010ad} or by projecting the static-light meson spins into the direction $\hat{\mathbf r}$ of the static source distance, by applying $\frac12(\Eins\pm i\gamma_5\pmb{\gamma} \cdot \hat{\mathbf{r}})$, and taking appropriate symmetric ($\Lambda_z=1$) or antisymmetric ($\Lambda_z=0$) spin combinations. These two approaches can be related to each other via a Fierz transformation. For our coarse lattice and the operator combinations that couple to total angular momentum $J=0$ we have performed this projection. For the other combinations and our fine lattice different representations will mix. The analyzed operators and the corresponding representations are listed in table \ref{tab:operators_potential}. We note that the operator combinations $\gamma_5 \times \gamma_5$ and $\Eins \times \Eins$ carry the same quantum numbers as well as the combinations $\gamma_5 \times \Eins$ and  $\gamma_5 \times \gamma_i\nabla_i$.
\TABULAR{|c|c|l|l|l|l|}{
\hline
$\mathcal{O} \; \otimes\; \mathcal{O}$ &  & \multicolumn{2}{l|} {Isospin: $I=1$}  & \multicolumn{2}{l|} {Isospin: $I=0$} \\
	\cline{3-6}
& $\!\!\Lambda_z$ & $r=0$: $J^{P}$ & $r>0$: $\Lambda_\eta^{(\sigma_\nu)}$ & $r=0$: $J^{PC}$ &$r>0$: $\Lambda_\eta^{(\sigma_\nu)}$\\
	\hline
	$\gamma_5 \;\otimes\; \gamma_5$, $\Eins \;\otimes\; \Eins$         & 0 & $0^{+}, 1^+$ & $\Sigma_g^{+}$ & $0^{++}, 1^{+-}$  &  $\Sigma_g^{+},\Sigma_u^-$ \\
	     & 1 & $1^+$        & $\Pi_g$                     & $1^{+-}$          &  $\Pi_u$ \\
	\hline
	$\gamma_5 \;\otimes\; \Eins$, $\gamma_5 \;\otimes\; \gamma_i\nabla_i$       & 0 & $0^{-}, 1^-$ & $\Sigma_u^{-}$ & $0^{-+}, 1^{--}$  &  $\Sigma_u^{-},\Sigma_g^+$ \\
	 & 1 & $1^-$        & $\Pi_u$                     & $1^{--}$          &  $\Pi_g$ \\
	\hline
	$\gamma_5 \;\otimes\; \left( \gamma_1\nabla_1 - \gamma_2\nabla_2 \right)$        & $/$ &  $1^{-},2^-$  &  $\Sigma_u^{+}, \Pi_u,\Delta_u$ & $1^{--}, 2^{-+}$   &  $\Sigma_g^{+}, \Pi_g, \Sigma_u^+$\\\hline
	$\gamma_i\nabla_i \;\otimes\; \left( \gamma_1\nabla_1 - \gamma_2\nabla_2 \right)$& $/$ &  $1^{+},2^+$  &  $\Sigma_g^{-}, \Pi_g,\Delta_g$ &  $1^{+-}, 2^{++}$  &  $\Sigma_u^{-}, \Pi_u, \Sigma_g^+$\\
	\hline}
{Operators and continuum representations for the meson-meson ($\mathcal{BB}$) and meson-antimeson ($\mathcal{B\overline{B}}$) potentials in the isosinglet and the isovector channel. \label{tab:operators_potential}}

\section{Results}
The eigenvalues $\lambda^{(k)}(t,t_0)$ of the generalized eigenvalue problem~\cite{Vari}, are fitted to one- and two-exponential ans\"atze, to obtain the $k$th mass. The appropriate values of $t_0$ and the fit ranges in $t$ are determined from monitoring the effective masses as described in~\cite{Bali:2010xa}. Let us first discuss $\mathcal{BB}$ meson systems. We define intermeson potentials as the differences between the meson-meson energy levels and the $r\rightarrow\infty$ two static-light meson limiting cases:
\begin{align}
V_{\!\mathcal{B}_1\mathcal{B}_2}(r)=E_{\!\mathcal{B}_1\mathcal{B}_2}(r) - 
	\left(m_{\!\mathcal{B}_1}+ m_{\!\mathcal{B}_2}\right)\quad\stackrel{r\rightarrow\infty}{\longrightarrow}\quad 0\,.
\end{align}
In the left panel of figure~\ref{fig:mp_diff_spin_isospin}, we display the ground state ($\Sigma_g^+$) and the first excited state ($\Sigma_g^{+\prime}$) of the $\gamma_5 \otimes \gamma_5$ operator as well as the $\Sigma_u^-$ ground state and the $\Sigma_u^{-\prime}$ first excited state of the $\gamma_5 \otimes \Eins$ operator, both in the $I=1,\Lambda_z=0$ channel. In the last case the lowest lying $r\rightarrow \infty$ combination of states would be a radially excited $\frac12^{+\prime}$ state ($\gamma_5)$ plus a $\frac12^-$ ($\Eins$) ground state. The next level would be the sum of $\frac12^+$ and $\frac12^{-\prime}$. It is not clear to which one of these states our creation operator has best overlap. In the figure we display both possibilities. The latter assignment would mean that in the excited state channel (like for the ground state) there is repulsion at intermediate distances.
 
In the right panel of figure~\ref{fig:mp_diff_spin_isospin} we show the $\gamma_5 \otimes \gamma_5$ operator in the different spin $\Lambda_z=0,1$ and isospin $I=0,1$ channels for the ground state. For short distances we observe attraction in all spin and isospin channels. In fact at very short distances we find attraction in all analyzed channels, see table~\ref{tab:operators_potential}, for ground and excited states. This may not be too surprising as this is expected from gluon exchange in the $3^*$ channel between the two static sources. When comparing the ground state of the $\gamma_5 \otimes \gamma_5$ operator for different spin and isospin channels we figure out that the $\Lambda_z=0$ channel is more attractive than the $\Lambda_z=1$ channel for isospin $I=0$. For isospin $I=1$ this pattern is reversed. In both cases the difference is of the order of $150$~MeV at a distance of $0.11$~fm. For the other spin-projected operator combination ($\Eins\otimes\Eins$, $\gamma_5\otimes\Eins$ and $\gamma_5\otimes\gamma_i\nabla_i$) we also find attractive forces of similar sizes for the $\Lambda_z=0,1$ and isospin $I=0$ ground states while for isospin $I=1$ the $\Lambda_z=1$ channel is more attractive than the $\Lambda_z=0$ channel.

On our coarse lattice we observe repulsive potentials at distances between $0.2$~fm and $0.45$~fm for the ground state of the $\gamma_5 \otimes \Eins$ operator in all spin and isospin channels. In addition we find repulsion in the $I=1,\Lambda_z=1$ channel for the $\Sigma_g^+$ ground state of the $\gamma_5 \otimes \gamma_5$ operator  at distances between $0.3$~fm and $0.45$~fm. In the case of the fine lattice we did not perform the $\Lambda_z$ projection so that here we cannot distinguish between $\Sigma$ and $\Pi$ states. In agreement with the coarse lattice results we obtain repulsion of $O(50\,\,\mathrm{MeV})$ in the ground states of the operator combinations $\gamma_5 \otimes \Eins, \gamma_5 \otimes \nabla_i\gamma_i\, (\Sigma_u^-)$ and $\gamma_5 \otimes \gamma_5\, (\Sigma_g^+)$ 
for isospin $I=1$ at intermediate distances $r>0.35$~fm. 
\FIGURE[t]
{\includegraphics[width=.49\textwidth,clip]{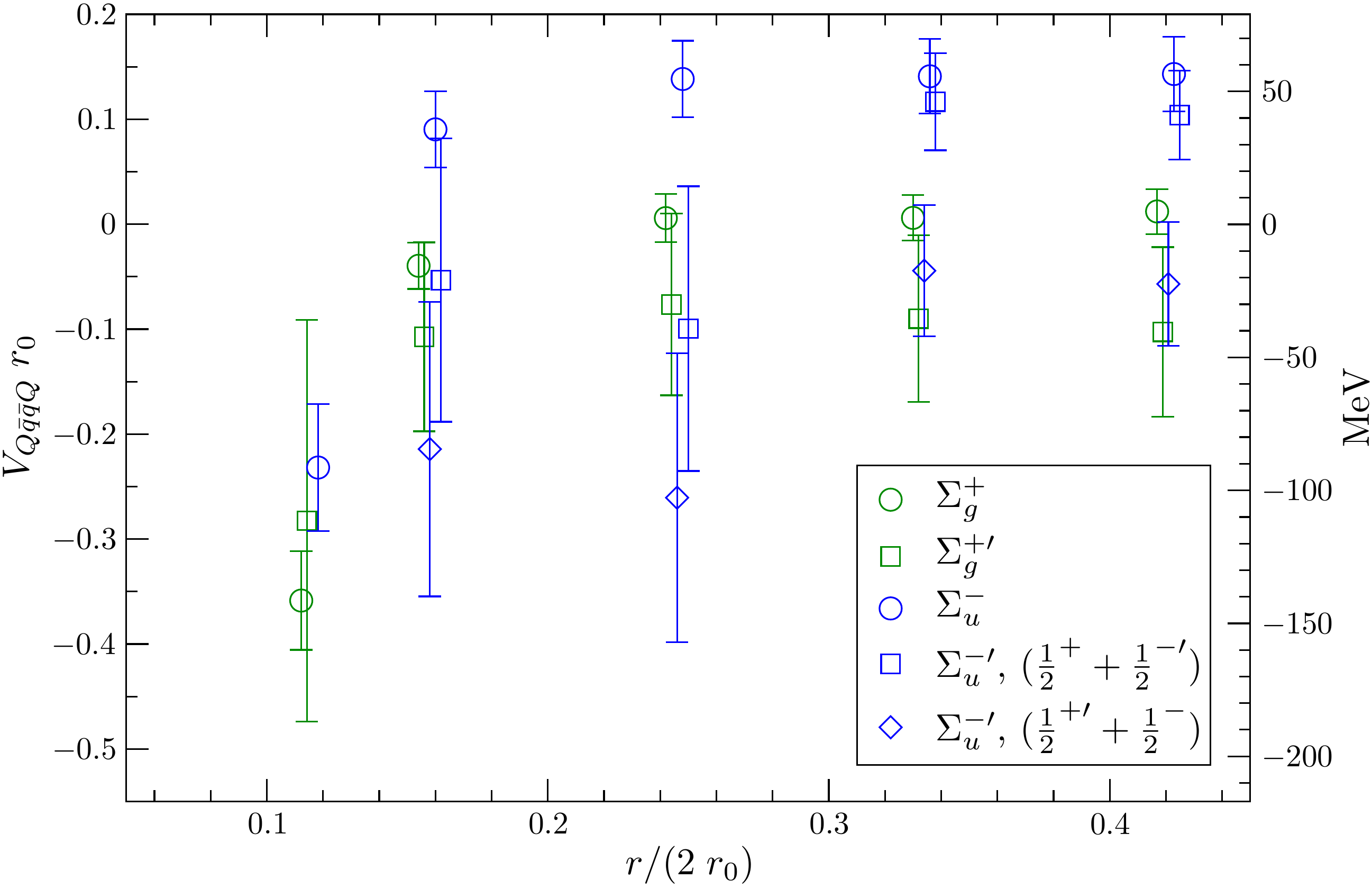}\hspace*{0.015\textwidth}
	\includegraphics[width=.49\textwidth,clip]{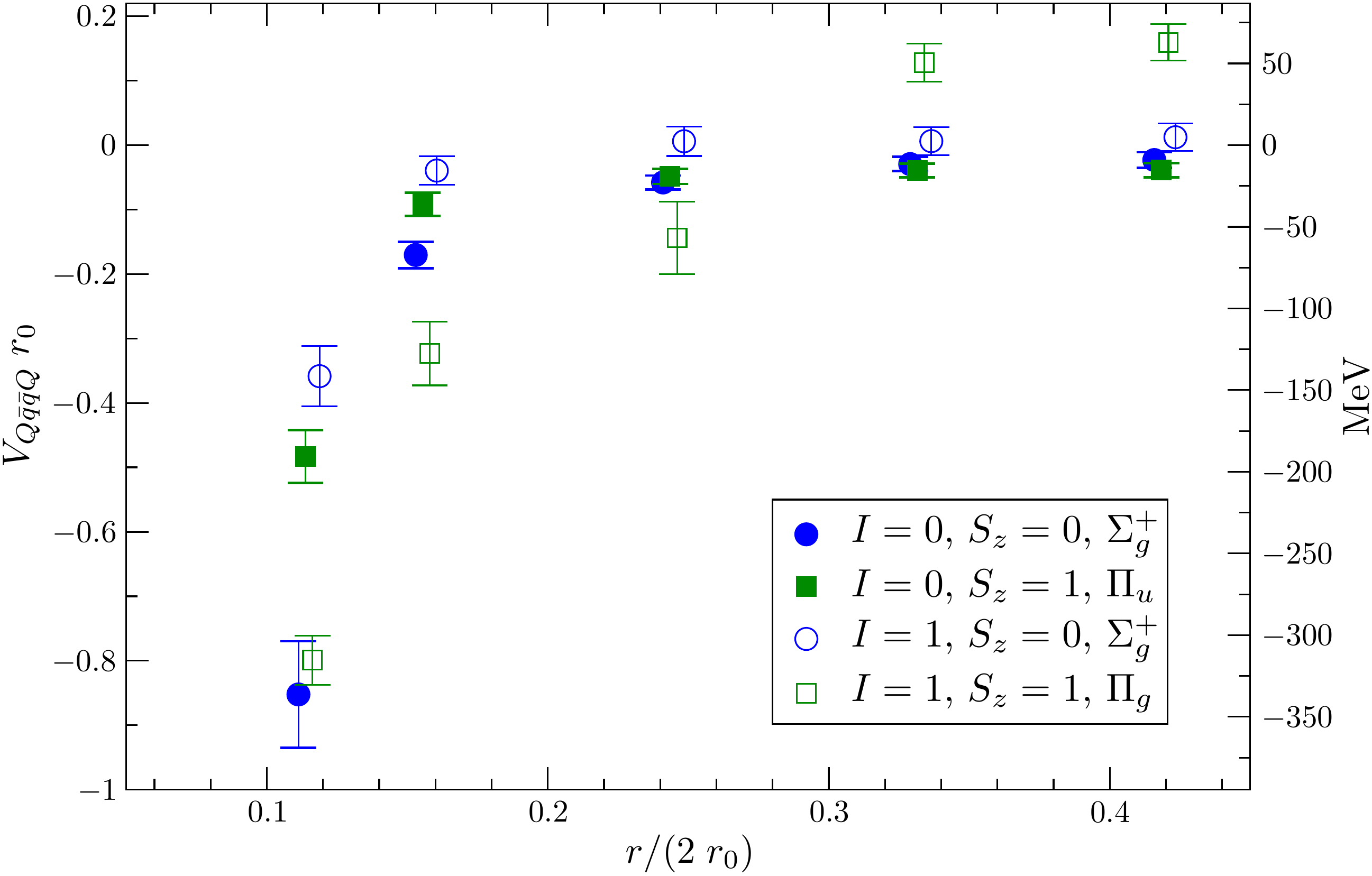}
\vspace*{-0.39cm}
\caption{Intermeson potentials $V(r)$ from our coarse lattice. The combinations $\gamma_5 \otimes \gamma_5$ ($\Sigma_g^+, \Sigma_g^{+\prime}$) and $\gamma_5 \otimes \Eins$ ($\Sigma_u^-, \Sigma_u^{-\prime}$) for $I=1,\Lambda_z=0$ are shown on the left hand side, the operator combination $\gamma_5 \otimes \gamma_5$ in different spin $\Lambda_z=0,1$ and isospin $I=0,1$ channels on the right hand side.
\label{fig:mp_diff_spin_isospin}}}
\FIGURE[t]
{\includegraphics[width=.49\textwidth,clip]{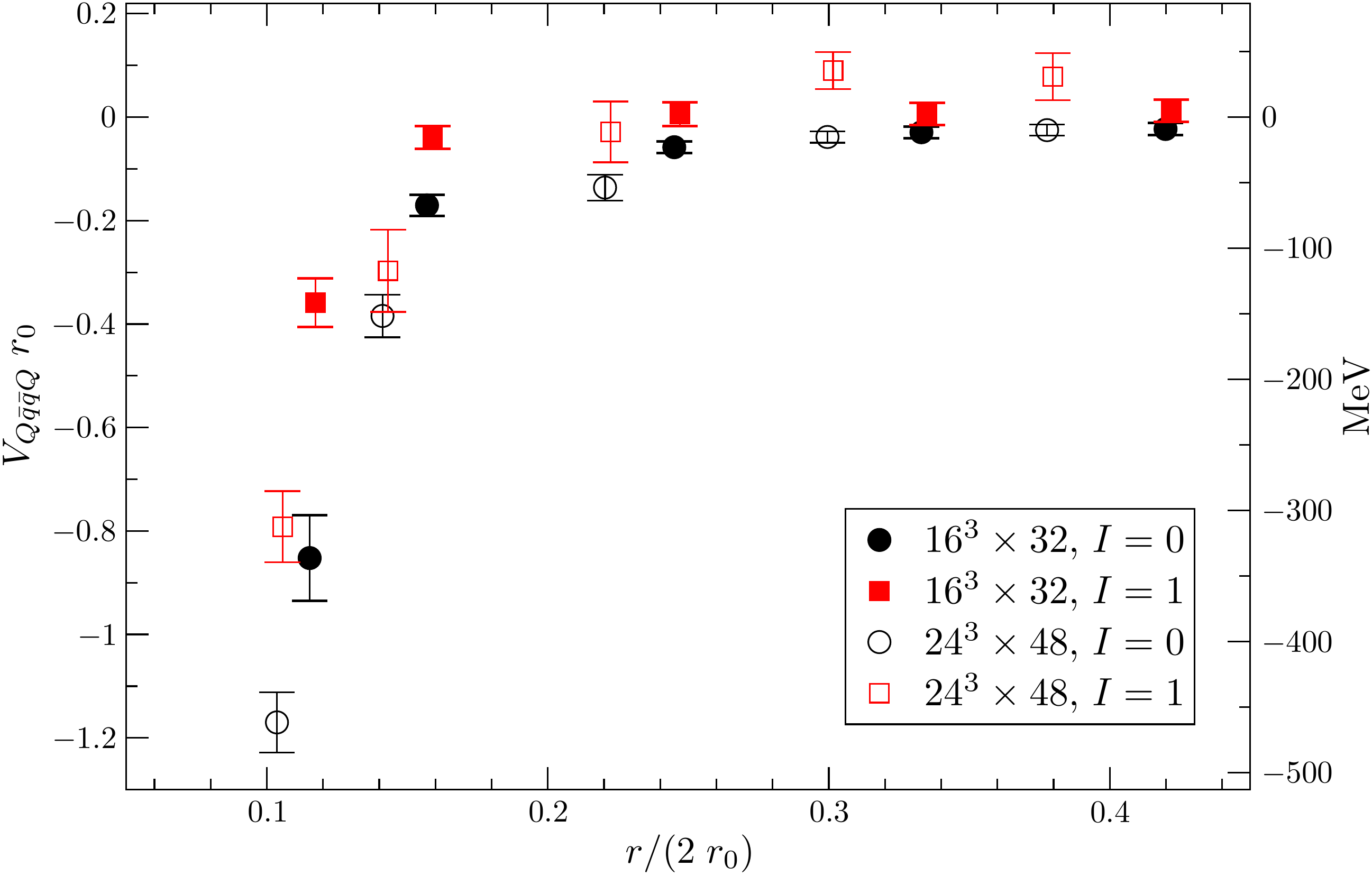}\hspace*{0.015\textwidth}
	\includegraphics[width=.49\textwidth,clip]{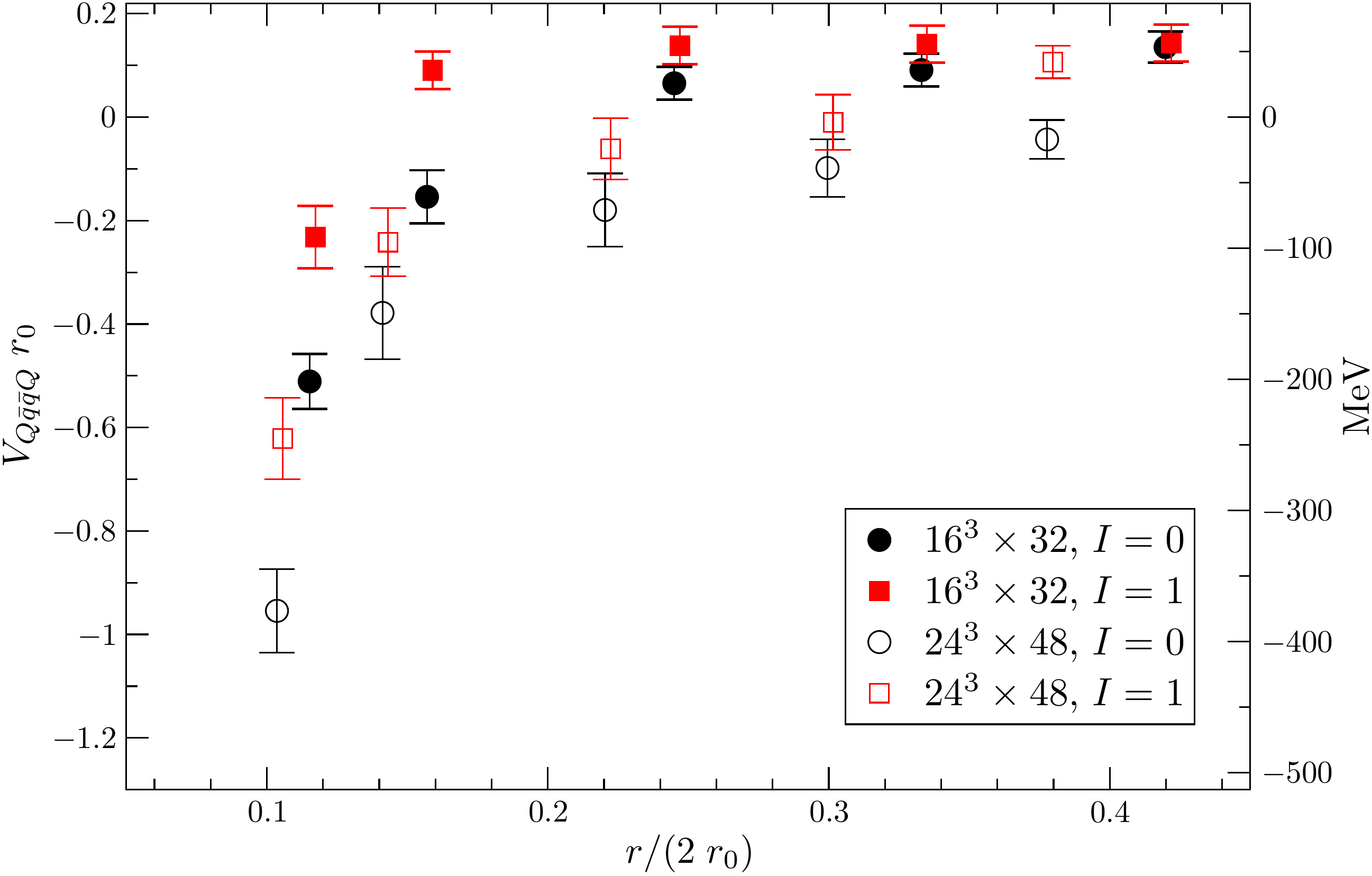}
\vspace*{-0.39cm}
\caption{Comparison of intermeson potentials $V(r)$ in different isospin channels between the results from our coarse and our fine lattice and heavy and light pion masses respectively. The combination $\gamma_5 \otimes \gamma_5$ ($\Sigma_g^+$) is shown on the left hand side, the combination $\gamma_5 \otimes \Eins$ ($\Sigma_u^-$) on the right hand side. \label{fig:mp_diff_quark_masses}}}

In figure~\ref{fig:mp_diff_quark_masses} we compare coarse and fine lattice results in the $\gamma_5 \otimes \gamma_5$ ($\Sigma_g^+$ ground state) and $\gamma_5 \otimes \Eins$ ($\Sigma_u^-$ ground state) channels. In the first channel we observe reasonable scaling while in the latter channel the fine lattice potentials appear to be more attractive. This may be related to the lighter pion mass resulting in a different Yukawa interaction.
\FIGURE[t]
{\includegraphics[width=.49\textwidth,clip]{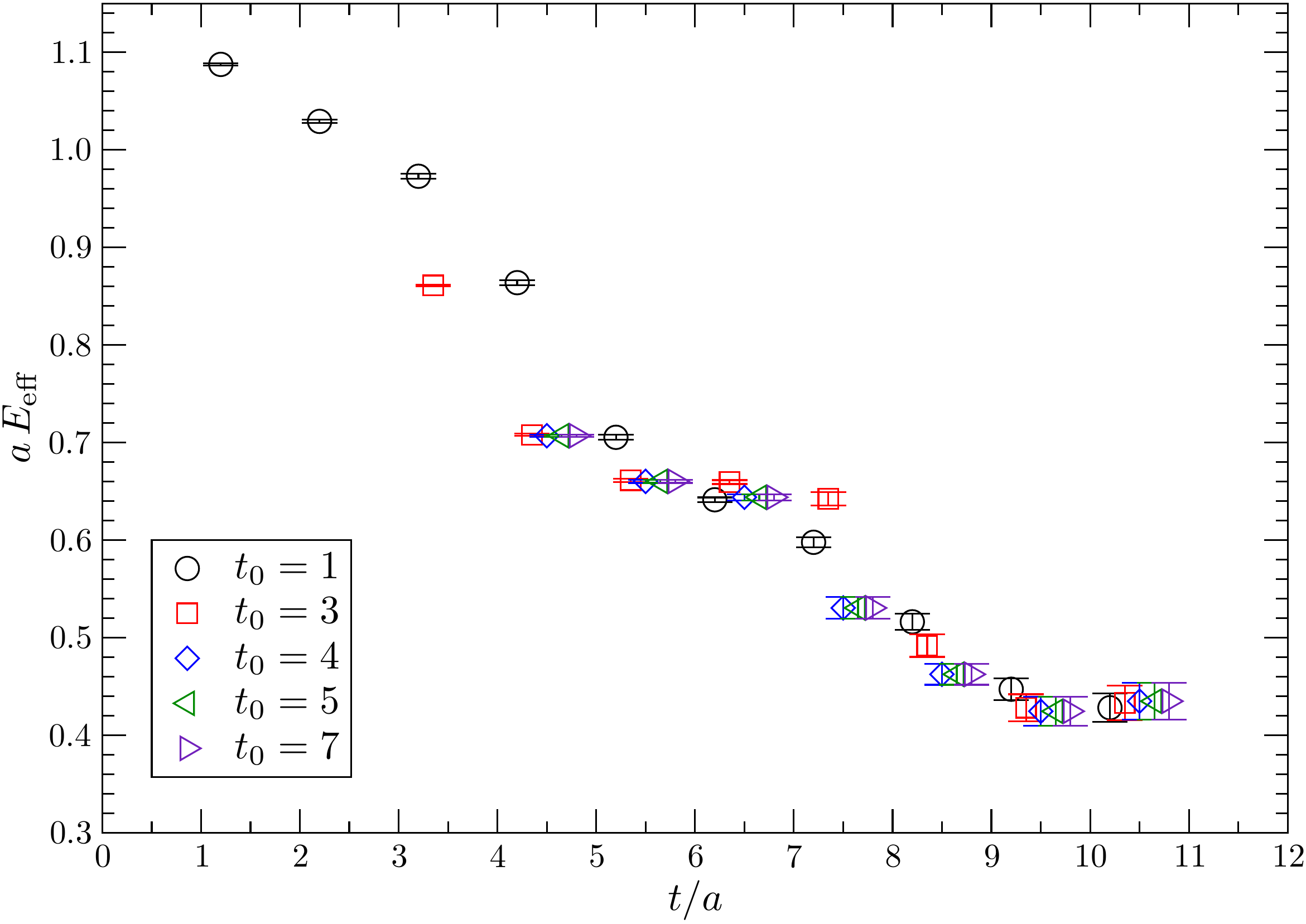}\hspace*{0.015\textwidth}
 \includegraphics[width=.49\textwidth,clip]{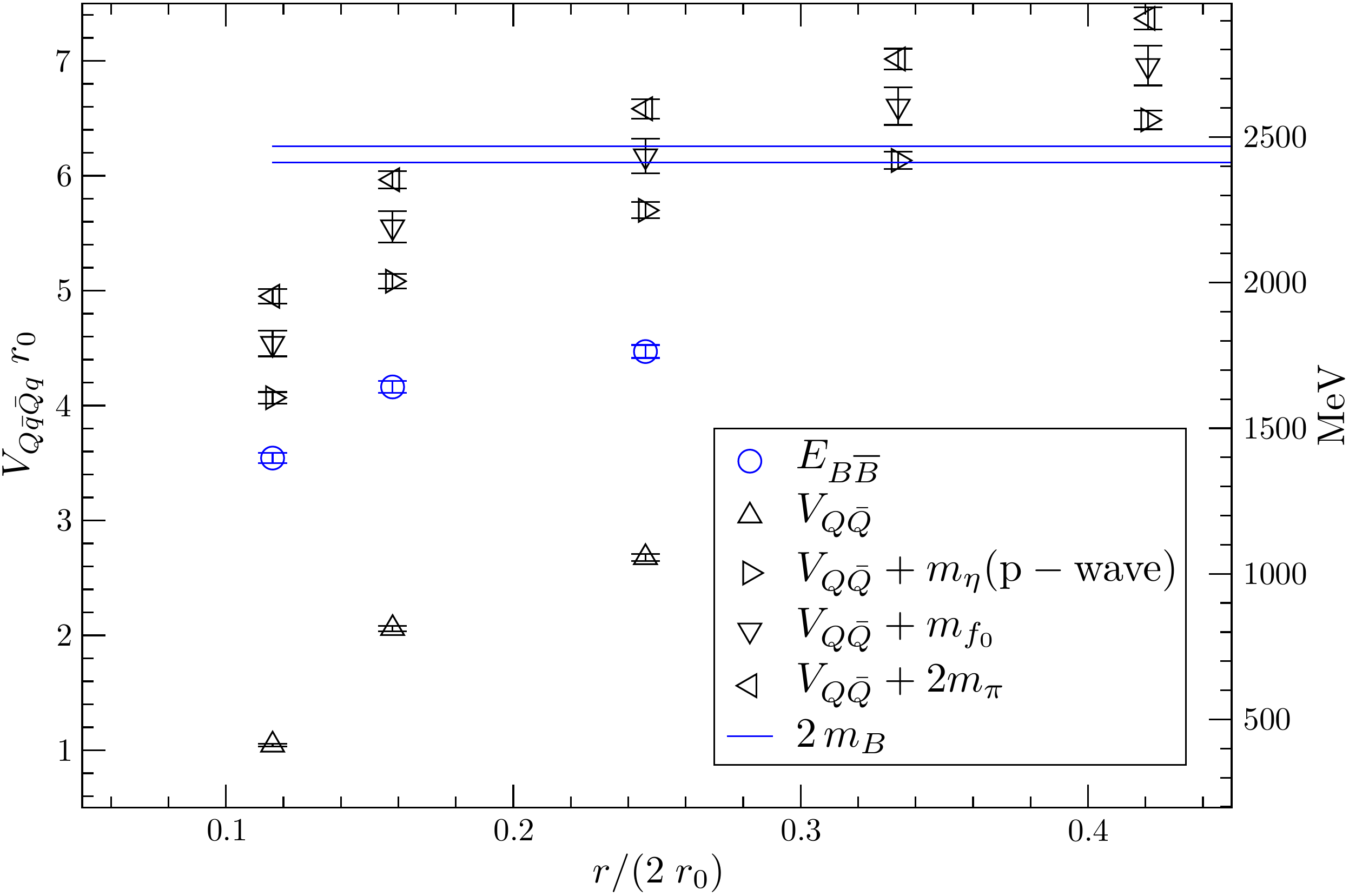}
\vspace*{-0.39cm}
\caption{Effective masses for the $\gamma_5 \otimes \gamma_5$ $\mathcal{B}\overline{\mathcal B}$ isoscalar $\Lambda_z=0$ ground state for different $t_0$ (left hand side). Masses for the $\gamma_5 \otimes \gamma_5$ $\mathcal{B}\overline{\mathcal B}$ ground state. The horizontal lines denote the sum of two static-light meson masses while the black symbols are the static $Q\overline{Q}$ potential plus various light meson masses (right hand side). \label{fig:mp_antipara_box}}} 

In figure~\ref{fig:mp_antipara_box} we display the $I=0$, $\Lambda_z=0$ ground state for the $\mathcal{B}\overline{\mathcal{B}}$ meson-antimeson case in the $\gamma_5 \otimes \gamma_5 \, (\Sigma_g^+$) channel. On the left hand side we see the effective ground state energy levels $E_{\mathrm{eff}}$ for different $t_0$ of the $\gamma_5 \otimes \gamma_5$ operator combination at a distance of $0.11$~fm. One finds a short plateau of poor quality in a range $t/a\in\{4,\ldots,8\}$ for $t_0>2a$. Then the effective energy level decreases again and forms another plateau from $t/a=8$ onwards. This can be explained by the observation that this state has the same quantum numbers as the $Q\overline{Q}$ static potential (and multiparticle states of the static potential plus a $P$ wave $\eta$ meson, the static potential plus 2 pions etc.). Our interpolator basis however, has very little overlap with these states. Therefore, we cannot easily disentangle the $Q\overline{Q}$ static potential and this background of multiparticle excitations from the lowest lying ${\mathcal B}\overline{\mathcal B}$ state that we are interested in. The ansatz to rewrite the correlator as $C_{ij}(t) =  C^{\mathcal{B\overline{B}}}_{ij}(t) + d_{ij}\cdot \exp(-V_{Q\overline{Q}}(t)\cdot t)$ and extract $C^{\mathcal{B\overline{B}}}_{ij}(t)$ failed as well. Thus, we tried to fit the ground state $\mathcal{B\overline{B}}$ mass without using the variational approach, but from our single correlation function that has the largest overlap with the $\mathcal{B\overline{B}}$ ground state. Masses could be extracted for separations $r<0.26$~fm. For larger distances the coupling to the $\mathcal{B\overline{B}}$ state could not be resolved. The result is displayed in the right panel of figure~\ref{fig:mp_antipara_box} (blue circles). The two horizontal lines correspond to twice the ground state mass of the $\frac12^+$ static-light meson, the expected $r\rightarrow\infty$ limit. At first sight there appear to be very substantial short distance attractive forces in this channel. Also the static potential $V_{Q\overline{Q}}$ is lying much lower and can be disentangled. However, states consisting of a static potential and a scalar $I=0$ particle will have the same quantum numbers. For our lattice parameters the $P$-wave pseudoscalar $\eta$ meson (that at our quark mass will have a similar mass to the pion) is the lowest such state, with masses of a $f_0$ meson as well as two pseudoscalars lying higher. We include these sums in the figure where we approximate $m_{\eta}$ by $m_{\pi}$ and $m_{f_0}$ by $m_{a_0}$. The ground state ${\mathcal B}\overline{\mathcal B}$ lies between these states and the static quark potential. So it is hard to decide whether we see a substantial attraction between the static-light meson-antimeson pair in this channel or bound states between the static quark potential and additional light mesons.

\section{Conclusions}
We investigated interactions between pairs of static-light mesons and found attraction for short distances in all spin and isospin channels. For distances of the order of $0.4$~fm some operator combinations yield repulsion, in particular the combination $\gamma_5 \otimes \Eins$. The interaction ranges are larger on the fine lattice with a smaller pion mass than on the coarse lattice. Meson-antimeson potentials are also very interesting with respect to charmonium threshold states~\cite{Brambilla:2010cs} ($D\overline{D}$ molecules or tetraquarks) but difficult to disentangle from mesons that are bound to the static potential (hadro-quarkonium~\cite{Dubynskiy:2008mq}). Analyzing $I=0, \Lambda_z=0$ ${\mathcal B}\overline{\mathcal B}$ states is a very challenging task since they couple directly to the $Q\overline{Q}$ static potential, e.g. in the case of the $\gamma_5 \otimes \gamma_5$ combination, or to the vacuum state, e.g. for the $\gamma_5 \otimes \Eins$ channel.

\acknowledgments
We thank Sara Collins, Christian Ehmann, Christian Hagen and Johannes Najjar for their help. We also thank our collaborators from the QCDSF Collaboration for generating the gauge ensemble. The computations were mainly performed on Regensburg's Athene HPC cluster. We thank Michael Hartung and other support staff. We acknowledge support from the GSI Hochschulprogramm (RSCHAE), the Deutsche Forschungsgemeinschaft (Sonderforschungsbereich/Transregio 55) and the European Union grant 238353, ITN STRONGnet.


\begin{thebibliography}{99}
\bibitem{Michael:1999nq}
  C.~Michael and P.~Pennanen  [UKQCD Collaboration],
  \emph{Two heavy-light mesons on a lattice},
  \emph{Phys.\ Rev.\  D}\ \textbf{60} (1999) 054012
  [\href{http://arxiv.org/abs/hep-lat/9901007}{\tt
arXiv:hep-lat/9901007}].
\bibitem{Pennanen:1999xi}
  P.~Pennanen, C.~Michael and A.~M.~Green  [UKQCD Collaboration],
  \emph{Interactions of heavy-light mesons},
  \emph{Nucl.\ Phys.\ Proc.\ Suppl.}\  \textbf{83} (2000) 200
  [\href{http://arxiv.org/abs/hep-lat/9908032}{\tt arXiv:hep-lat/9908032}].
\bibitem{Detmold:2007wk}
  W.~Detmold, K.~Orginos and M.~J.~Savage,
  \emph{$BB$ potentials in quenched lattice QCD},
  \emph{Phys.\ Rev.\  D}\ \textbf{76} (2007) 114503
  [\href{http://arxiv.org/abs/hep-lat/0703009}{\tt arXiv:hep-lat/0703009}].
\bibitem{Sesam} 
	G.~S.~Bali, H.~Neff, T.~D\"ussel, T.~Lippert and K.~Schilling, 
 	\emph{Observation of string breaking in QCD}, 
	\emph{Phys.\ Rev.\ D}\ \textbf{71} (2005) 114513 
	[\href{http://arxiv.org/abs/hep-lat/0505012}{{\tt arXiv:hep-lat/0505012}}].
\bibitem{Wagner:2010ad}
  M.~Wagner  [ETM Collaboration],
  \emph{Forces between static-light mesons},
  \href{http://arxiv.org/abs/1008.1538}{\tt arXiv:1008.1538 [hep-lat]}.
\bibitem{Wagner:2011ev}
  M.~Wagner  [ETM Collaboration],
  \emph{Static-static-light-light tetraquarks in lattice QCD},
  \href{http://arxiv.org/abs/1103.5147}{\tt arXiv:1103.5147 [hep-lat]}.
\bibitem{Bali:2010xa}
	G.~S.~Bali and M.~Hetzenegger [QCDSF Collaboration],
	\emph{Static-light meson-meson potentials},
	\emph{PoS} \textbf{LAT2010} 142
	\href{http://arxiv.org/abs/1011.0571}{\tt arXiv:1011.0571 [hep-lat]}.
\bibitem{AliKhan:2003br}
  A.~Ali Khan \textit{et al.}  [QCDSF Collaboration],
  \emph{Accelerating the hybrid Monte Carlo algorithm},
  \emph{Phys.\ Lett.\  B}\ \textbf{564} (2003) 235
[\href{http://arxiv.org/abs/hep-lat/0303026}{\tt arXiv:hep-lat/0303026}].
\bibitem{Edwards:2004sx}
R.~G.~Edwards and B.~Jo\'o,
\emph{The Chroma software system for Lattice
      QCD}, \emph{Nucl.\ Phys.\ Proc.\ Suppl.}\ \textbf{140} (2005) 832
[\href{http://arxiv.org/abs/hep-lat/0409003}{{\tt hep-lat/0409003}}];
\bibitem{Vari}
	C.~Michael, 
 	\emph{Adjoint sources in Lattice Gauge Theory},
	\emph{Nucl.\ Phys.\ B}\ \textbf{259} (1985) 58.
\bibitem{Wuppertalsmear}
	S.~G\"usken \textit{et al.},
	\emph{Non-singlet axial vector couplings of the baryon octet in lattice QCD}, 
	\emph{Phys.\ Lett.\ B}\ \textbf{227} (1989) 266.
\bibitem{Michael:1998sg}
	C.~Michael and J.~Peisa, 
 	\emph{Maximal variance reduction for stochastic propagators with applications to the static quark spectrum},
	\emph{Phys.\ Rev.\ D}\ \textbf{58} (1998) 034506
[\href{http://arxiv.org/abs/hep-lat/9802015}{\tt arXiv:hep-lat/9802015}].
\bibitem{Brambilla:2010cs}
	N.~Brambilla \textit{et al.},
	\emph{Heavy quarkonium: progress, puzzles, and opportunities},
	\href{http://arxiv.org/abs/arXiv:1010.5827}{\tt arXiv:1010.5827 [hep-ph]}.
\bibitem{Dubynskiy:2008mq}
  S.~Dubynskiy and M.~B.~Voloshin,
  \emph{Hadro-charmonium},
  \emph{Phys.\ Lett.\  B}\ \textbf{666} (2008) 344
[\href{http://arxiv.org/abs/0803.2224}{\tt arXiv:0803.2224 [hep-ph]}].
\end{thebibliography}
\end{document}